# Quantum Particle Swarm Optimization for Electromagnetics

Said Mikki and Ahmed A. Kishk
Center of Applied Electromagnetic Systems Research,
Department of Electrical Engineering, University of Mississippi,
University, MS 38677, USA

*Abstract* – **A new particle swarm optimization (PSO) technique for electromagnetic applications is proposed. The method is based on quantum mechanics rather than the Newtonian rules assumed in all previous versions of PSO, which we refer to as *classical* PSO. A general procedure is suggested to derive many different versions of the quantum PSO algorithm (QPSO). The QPSO is applied first to linear array antenna synthesis, which is one of the standard problems used by antenna engineers. The performance of the QPSO is compared against an improved version of the classical PSO. The new algorithm outperforms the classical one most of the time in convergence speed and achieves better levels for the cost function. As another application, the algorithm is used to find a set of infinitesimal dipoles that produces the same near and far fields of a circular dielectric resonator antenna (DRA). In addition, the QPSO method is employed to find an equivalent circuit model for the DRA that can be used to predict some interesting parameters like the Q-factor. The QPSO contains only one control parameter that can be tuned easily by trial and error or by suggested simple linear variation. Based on our understanding of the physical background of the method, various explanations of the theoretical aspects of the algorithm are presented.**

## I. Introduction

The evolutionary PSO is a global search strategy that can handle efficiently arbitrary optimization problems. In 1995, Kennedy and Eberhart introduced the PSO method for the first time [1]. Later, it received a considerable attention and proved to be capable of tackling difficult optimization problems. The basic idea of the PSO is to mimic the social interactions between members of biological swarms. One of the good examples illustrating the concept is the analogy with the swarm of bees. Bees (solution candidates) are allowed to fly in a specified field, looking for food. It is believed that after certain time (generations; iterations) all bees will gather around the highest concentration of food in the field (global optimum). At every generation, each bee updates its current location by employing information about the local and global "bests", achieved so far, received from all other bees. Such social interactions and continuous velocity update will guarantee arriving to the global optimum. The method has received considerable attention by the electromagnetic community because of its simplicity and high capability of searching for the global optimum of hard optimization problems. The classical PSO method was recently applied to electromagnetic problems [2]-[4], [11] and proved to be very competitive compared with other well-established evolutionary computing techniques, such as the genetic algorithm [4].

A quantum-inspired version of the PSO algorithm (QPSO) was proposed very recently [5],[6]. The QPSO algorithm permits all particles to have a quantum behavior instead of the classical Newtonian dynamics that was assumed so far in all versions of the PSO. Thus, instead of the Newtonian random walk, some sort of "quantum motion" is imposed in the search process. When the QPSO is tested against a set of benchmarking functions, it demonstrated superior performance as compared to the classical PSO but under the condition of large population sizes [5]. One of the most attractive features of the new algorithm is the reduced number of control parameters. Strictly speaking, there is only one parameter required to be tuned in the QPSO.

In this article, a generalized framework that allows the user to derive many versions of the QPSO method is proposed where distinct potential well types are presented. A physical interpretation of the algorithm is presented through the discussion of different possible potential wells. Based on our understanding of the physical roots of the new strategy, we propose guidelines to control the tuning parameter of the algorithm. We first introduce the QPSO by illustrating its application to linear array antenna synthesis problems. By conducting several computer experiments and comparing the performances of the two algorithms, the QPSO was found to outperform the classical PSO. Then, the new algorithm is used to study the application of antenna modeling using a set of infinitesimal dipoles. By formulating the modeling of circular dielectric resonator antenna (DRA) as an optimization problem, the QPSO algorithm was able to find a model of 10 dipoles, predicting accurately both the near and far fields. Finally, the method is employed in finding an equivalent circuit to study the resonance of the antenna.

## II. The Classical PSO

It will be very instructive to review first the basics of the PSO method in order to introduce the quantum version. Assume that our problem is $N$-dimensional. The following *position* vector represents the time evolution of a set of $M$ particles

$$\mathbf{x}^m(t) = \begin{bmatrix} x_1(t) & x_2(t) & . & . & . & x_N(t) \end{bmatrix}^T \tag{1}$$

The velocity vector is given by

$$\mathbf{v}^m(t) = \begin{bmatrix} v_1(t) & v_2(t) & . & . & . & v_N(t) \end{bmatrix}^T \tag{2}$$

where $T$ is the transpose operator. The superscript $m$ is an index ranging from 1 to $M$. The core idea of the classical PSO algorithm is the exchange of information about the global and local best values. This can be done by the following two equations

$$\mathbf{v}^m(t+\Delta t) = w\mathbf{v}^m(t) + c_1\left[\mathbf{\Phi}_1\right]\left(\mathbf{P}_L^m - \mathbf{x}^m(t)\right) + \mathrm{c}_2\left[\mathbf{\Phi}_2\right]\left(\mathbf{P}_g - \mathbf{x}^m(t)\right) \tag{3}$$

$$\mathbf{x}^m(t+\Delta t) = \mathbf{x}^m(t) + \Delta t\mathbf{v}^m(t) \tag{4}$$

where $\Delta t$ is the time step, $\mathbf{P}_L^m$ is the local best vector of the $m$th particle, and $\mathbf{P}_g$ is the global best vector. The elements of the two diagonal matrices

$\left[\mathbf{\Phi}_i\right] = \text{diag}(\varphi_1^i, \varphi_2^i, ......, \varphi_N^i)$ , $i = 1,2$, are a set of statistically independent random variables uniformly distributed between 0 and 1, the parameter $w$ is the inertia factor, and $c_1$ and $c_2$ are the cognitive and social factors, respectively. Equation (3) can be further simplified to be in the form

$$\mathbf{v}^m(t+\Delta t) = w\mathbf{v}^m(t) + \left[\mathbf{\Phi}\right]\left(\mathbf{P}^m - \mathbf{x}^m(t)\right) \tag{5}$$

where

$$\begin{aligned}\mathbf{P}^m = c_1 \,\text{diag}&\left(\frac{\varphi_1^1}{c_1\varphi_1^1 + c_2\varphi_1^2}, \frac{\varphi_2^1}{c_1\varphi_2^1 + c_2\varphi_2^2}, ...., \frac{\varphi_N^1}{c_1\varphi_N^1 + c_2\varphi_N^2}\right)\mathbf{P}_L^m \\ &+ c_2 \,diag\left(\frac{\varphi_1^2}{c_1\varphi_1^1 + c_2\varphi_1^2}, \frac{\varphi_2^2}{c_1\varphi_2^1 + c_2\varphi_2^2}, ...., \frac{\varphi_N^2}{c_1\varphi_N^1 + c_2\varphi_N^2}\right)\mathbf{P}_g\end{aligned} \tag{6}$$

$$\left[\mathbf{\Phi}\right] = \text{diag}\left(c_1\varphi_1^1 + c_2\varphi_1^2, c_1\varphi_2^1 + c_2\varphi_2^2, ...., c_1\varphi_N^1 + c_2\varphi_N^2\right) \tag{7}$$

It can be shown in [7] that for the PSO algorithm to converge, all particles must approach the location $\mathbf{P}^m$ given by (6). Such convergence can be obtained by a proper tuning of the cognitive and social parameters of the algorithm [7]. Also, to prevent the explosion of the particles in the classical PSO, a maximum velocity $V_{\max}$ is introduced in each dimension to confine swarm members inside the boundary walls of the domain of interest. Usually $V_{\max}$ is set to be the maximum of the dynamic range of the corresponding coordinate [3].

## III. Quantum Formulation of the Swarm Dynamics

The QPSO algorithm allows all particles to move under quantum-mechanical rules rather than the classical Newtonian random motion. In the classical environment, all bees are flying toward the "optimum" location defined by $\mathbf{P}^m$ in (6). The particles are then *attracted* to this location through the optimization process. Such attraction leads to the global optimum. From equation (6) it is easy to see that this location $\mathbf{P}^m$ is nothing but a random average of the global and local bests of the particles of the swarm. In quantum mechanics, the governing equation is the general time-dependent Schrödinger equation

$$j\hbar\frac{\partial}{\partial t}\Psi(\mathbf{r},t) = \hat{H}(\mathbf{r})\Psi(\mathbf{r},t) \tag{8}$$

where $\hat{H}$ is a time-independent Hamiltonian operator given by

$$\hat{H}(\mathbf{r}) = -\frac{\hbar^2}{2m}\nabla^2 + V(\mathbf{r}) \tag{9}$$

where $\hbar$ is Planck's constant, $m$ is the mass of the particle, and $V(\mathbf{r})$ is the potential energy distribution. In the Schrödinger equation, the unknown is the wave function $\Psi(\mathbf{r},t)$, which has no direct physical meaning. However, its amplitude squared is a probability measure for the particle's motion. By imposing the following normalization condition we can justify such a measure

$$\iiint |\Psi(\mathbf{r},t)|^2 dx\,dy\,dz = 1.0 \tag{10}$$

where the integration is performed over the entire space.

In the present version of the QPSO algorithm, we apply an attractive potential field that will eventually pull all particles to the location defined by (6) [5]. In quantum mechanics, this implies that the potential field will generate *bound states* [8].

To simplify the formulation, assume that we have the 1-D problem consisting of one particle moving in the dimension *r*. Let $r = x - p$, where $p$ is the average best given by (6). For particles to converge according to the QPSO algorithm, $r$ should approach zero. Therefore, we need to apply an *attractive* potential field centered at the zero. In principle, any potential *well* can work but the simplest one is the *delta-well* given by [8]-[9]

$$V(r) = -\gamma\,\delta(r) \tag{11}$$

where $\gamma$ is a positive number proportional to the "depth" of the potential well. The depth is infinite at the origin and zero elsewhere. Thus, the delta potential well is an idealized realization of an infinitely strong attractive potential field that works at a single location.

Assuming the principle of separation of variables, we separate the time dependence of the wave function from the spatial dependence. Substituting the separated form into (8) we get [8]

$$\Psi(r,t) = \psi(r)e^{-jEt/\hbar} \tag{12}$$

where $E$ represents the energy of the particle. The wave envelope $\psi(r)$ can be found by solving the following time-*independent* Schrödinger equation

$$\left\{-\frac{\hbar^2}{2m}\nabla^2 + V(r)\right\}\psi = E\,\psi \tag{13}$$

Thus, the solutions are time-independent. Such solutions are called *stationary* states. Non-stationary states can be formed by superposition of the eigen-solutions obtained from the time-independent Schrödinger equation. However, in this article we consider pure eigen-states only, and more specifically, *bound states*.

## IV. The Choice of the Potential Well Distribution

The next critical step in deriving the QPSO algorithm is the choice of a suitable attractive potential field that can guarantee bound states for the particles moving in the quantum environment. In the following paragraphs we list some of the possible choices.

### A. The Delta Potential Well

Here, we apply the potential distribution defined in (11). This choice leads to the simplest analytical solution possible for Schrödinger equation. It can be shown that under this potential field we get the expression of the probability density function (*pdf*) as shown in Table I [8], [9]. $L = \hbar^2 / m\gamma$ is a parameter called the *characteristic length* of the potential well. Thus, for a given system of particles of mass *m* we can directly control the characteristic length by varying the "depth" of the well $\gamma$.

### B. The Harmonic Oscillator

Another potential distribution, which is very common in quantum mechanics, is the harmonic oscillator potential well given by

$$V(r) = \frac{1}{2} k r^2 \tag{14}$$

where *k* is a parameter defining the well "depth" or "strength". Again, this problem has the following well-known analytical solution [8]

$$\psi_n(r) = \left( \frac{\alpha}{2^n n! \pi^{\frac{1}{2}}} \right)^{\frac{1}{2}} H_n(\alpha r) e^{-\frac{1}{2}\alpha^2 r^2} \tag{15}$$

where $\alpha = \left( mk / \hbar^2 \right)^{1/4}$ and $H_n$ is the Hermite polynomial. Equation (15) shows that multiple possible eigen-states exits in this system, each with integer index *n*. However, we may simplify the problem considerably by assuming that only the lowest possible mode (the ground state $n = 0$) is available. In this case, the Gaussian probability distribution can be obtained as illustrated in Table I. Again, the characteristic length of this well can be seen to be $\sqrt{\pi}/\alpha$, a quantity that is directly controlled by the strength of the well *k*.

### C. The Square Potential Well

A simple non-confining potential distribution in quantum mechanics is the square well [8] defined by

$$V(r)=\begin{cases}0 & , \ |r|\leq W/2 \\ V_1 & , \ \text{else}\end{cases} \quad (16)$$

It represents *energy walls* that are used to confine all particles with energy less than $V_1$ inside the walls located between the points $r=\pm W/2$. Although this potential well is more physically realizable, if compared with the "pathological" delta function well, the solution of Schrödinger equation becomes much more demanding. The state function can be obtained in analytical form, but many modes will be available (excited), depending on the relation between the depth of the well $V_1$ and its width $W$ [9]. For simplicity, we allow the existence of one mode only, which is the first or the ground mode (even mode) given by the expression shown in Table I. Here $a$, $b$, $\xi$, and $\eta$ are constants that depend on the choice of $\gamma$ in the relation $V_1 W=\gamma$. Note that by fixing the product of the depth $V_1$ and the width $W$ to a constant value $\gamma$, we force one mode only to exist [9]. Thus, only the parameter $W$ is needed in the description of the state of the particle.

### D. Other Potential Well Distributions

In quantum mechanics, other attractive potential field distributions that can be used in the implementation of the QPSO are possible. For example, we have the Yukawa potential $V_{\text{Yukawa}}(r)=-V_0 e^{-\mu r}/(\mu r)$; the Gaussian potential $V_{\text{Gauusian}}(r)=-V_0 e^{-(\beta r)^2}$; and the Woods-Saxon potential $V_{\text{W-S}}(r)=-V_0\left\{1+\exp\left[\left(r-R\right)/a\right]\right\}^{-1}$ [8]. However, none of these potentials leads to a tractable analytical solution. As we will see in the following sections, to derive a simple and efficient QPSO algorithm, it is very important to have analytical forms for the state function to be easily inverted.

## V. The Collapse of the Wave Function

So far we have assumed that all particles behave under the influence of Schrödinger equation solved for several potential well distributions. According to Heisenberg principle of uncertainty [8]-[9], it is impossible to define simultaneously both the position and the velocity of a particle with arbitrary accurate precision. Thus, in the QPSO we expect that no velocity term will exist in the basic update equations, in contrast to the case with the classical PSO depicted by equations (3) and (4). Instead, we need to "measure" the location of the particle. This is the fundamental problem in quantum mechanics. Specifically, measuring devices obey Newtonian laws while the particle itself follows the quantum rules. To interface between the two different worlds, one needs to "collapse" the wave function of a moving particle into the localized space of the measurement. This localization process can be easily accomplished through the following Monte Carlo simulation procedure: 1) generate a random variable uniformly distributed in the local space where the measurement is done, 2) equate the uniform distribution to the true probability distribution estimated by the quantum mechanics, and 3) solve for the position $r$ in terms of the random variables assumed.

By Generating a random variable $u$ uniformly distributed between 0 and 1, steps (2) and (3) above lead to the update equations illustrated in Table I. Notice that the localization of the probability density function for the square well case requires first to realize that the "natural" characteristic length is the width of the square well itself, $W$.

## VI. Selecting the Parameters of the Algorithm

The fundamental condition of convergence in any QPSO algorithm is given by

$$\lim_{t\to\infty} L(t) = 0 \tag{17}$$

where $L(t)$ is the time-dependent characteristic length. This can be directly inferred from the update equations in Table I where convergence is clearly understood to be the case with $\mathbf{x}^m \to \mathbf{P}^m$ for all $m$. Thus, we need to enforce a time evolving parameter $L(t)$ such that all particles will eventually arrive to the desired location. To guarantee, on the average, that the next particle will converge, we need the value of $|r_{k+1}|$ at iteration $k$ to be closer to zero. A suitable probabilistic translation for this statement is given by

$$\int_{-\infty}^{|r_k|} Q_{k+1}(r)dr > 0.75 \tag{18}$$

where $Q_{k+1}(r)$ is the probability density function of the particle at the $(k+1)th$ iteration. The reader should notice that convergence obtained under condition (18) is guaranteed only in the probabilistic sense. Some particle may diverge, but on the average the algorithm should converge. This relation follows from the fact that the integration from $-\infty$ to 0 will produce 0.5, so the "remaining" probability of 0.5 should be divided between the decision that the particle location in the $(k+1)th$ iteration will be either to the left or to the right of $|r_k|$. The condition of (18) leads to the desired situation where we have a higher probability of approaching the origin.

By solving (18) for the delta potential well, we get

$$L_{k+1} < \frac{1}{\ln\sqrt{2}}|x_k - p| \tag{19}$$

Thus, we may choose

$$L_{k+1} = \frac{1}{g(\ln\sqrt{2})}|x_k - p| \tag{20}$$

The condition in (19) is automatically satisfied when

$$g > 1 \tag{21}$$

This leads to the Quantum Delta PSO (QDPSO).

The derivation of the corresponding formulas for the harmonic oscillator and the square potential wells requires more effort because of the difficulty in carrying out the integration of (18). For the harmonic oscillator, by numerically solving an equation involving the error function, we obtained

$$\alpha = g \frac{0.47694}{|x_k - p|} \tag{22}$$

This leads to the Quantum Harmonic Oscillator PSO (QOPSO).

For the square well, we obtained the following through a much longer calculations

$$W = \frac{0.6574}{g} |x_k - p| \tag{23}$$

This leads to the Quantum Square-Well PSO (QSPSO). Notice that condition (21) is required in (22) and (23).

Table I summarizes the *pdf*s and the update equations of the various potential wells derived above.

**Table I** Summary of the probability density functions (*pdf*) and update equations of the delta, harmonic, and square potential wells

| Potential Well\Derived quantity | *pdf* | Update Equation |
|---|---|---|
| Delta Potential Well | $Q(r) = \frac{1}{L} e^{-2\frac{\|r\|}{L}}$ | $x_{k+1} = p \pm \frac{\ln(1/u)}{2g \ln\sqrt{2}} \|x_k - p\|$ |
| Harmonic Oscillator | $Q(r) = \frac{\alpha}{\sqrt{\pi}} e^{-\alpha^2 r^2}$ | $x_{k+1} = p \pm \frac{1}{0.47694g} \sqrt{\ln\left(\frac{1}{u}\right)} \|x_k - p\|$ |
| Square Well | $Q(r) = \begin{cases} \frac{a}{W}\cos^2(\frac{\xi}{W} r) & , \|r\| \le W/2 \\ \frac{b}{W} e^{-\frac{\eta}{W} r} & , r \ge W/2 \\ \frac{b}{W} e^{\frac{\eta}{W} r} & , r \le -W/2 \end{cases}$ | $x_{k+1} = p + \frac{0.6574}{\xi g} \cos^{-1}\left[\pm\sqrt{u}\right] \|x_k - p\|$ |

## VII. The QPSO Algorithm

### A. The Algorithm

Now, it is possible to summarize the proposed QPSO algorithm as follows:

1. Choose a suitable attractive potential well centered around the vector **P** given by equation (6) (Delta well, harmonic oscillator, square well, combinations of the previous wells, .., etc). Solve Schrödinger equation to get the wave function, and then the probability density function of the position of the particle.

2. Use Monte Carlo simulation – or any other measurement method – to collapse the wave function into the desired region. The result of this step is an equation in the form

$$\mathbf{x} = \mathbf{P} + f\ (L, \pm\mathbf{u}) \tag{24}$$

   where $\boldsymbol{u}$ is uniformly distributed random variables; $L = L\left(g, u, \left|x^{m} - p\right|\right)$ is the characteristic length, which is a function of $g, u,$ and $\left|x^{m} - p\right|$. The functional form *f* is obtained by the inversion of the probability density function.

3. Apply the pseudo-code shown in Fig. 1.

Note that the cognitive and social parameters do not appear in the QPSO algorithm. This is clear for the case of $c_1 = c_2$ where from equation (6) these factors cancel. Most of the research papers on the classical PSO shows that these two factors are better to be chosen equal [3], making this version of the QPSO justified. Values for the cognitive and social coefficients that are not equal to each other were used, but poor performance was obtained for all objective functions chosen. Thus, the generalized QPSO shown above contains one control parameter only, *g*, which is directly related to the characteristic length of the potential well. This makes the QPSO more attractive for electromagnetic applications compared with the classical PSO that requires extra parameters to be tuned for each application.

**B. Physical Interpretation**

Fig. 2 shows plots for probability density functions for the three potential wells derived above. The functions are normalized, in amplitude and position, to the characteristic length of the well in order to understand the qualitative differences between them. It is clear that the harmonic oscillator leads to the "tightest" distribution around the origin, and thus more particles are likely to be close to *p*, resulting in faster convergence according to (17). However, in a phenomenon similar to the well-known tunneling effect in quantum mechanics [8], some particles with very small probabilities, however still non-vanishing, are allowed to explore regions far away from the origin (the center of the potential well). This is in direct contrast with the rules of the classical mechanics where the attractive field will eventually pull *all* particles that do not have sufficient energies to escape. Thus, we expect that the QPSO will have a much stronger "insight" on the optimization space since it can "spy" on far regions in the domain of interest all the time. The "spies" are very few particles existing in the tail region of the probability density functions.

On the other hand, it is also interesting to observe how different choices of the potential distributions have led to different *probability* dynamics as described by the probability density functions $Q$ in Table I. The physical interpretation of the wave (state) function $\Psi$ in quantum mechanics, as a probability amplitude distribution describing the location of the particle, means that the new QPSO algorithm corresponds to a new probability law for the position vectors of the competing particles. For example, from the corresponding equation in Table I we see that the quantum delta (QDPSO) algorithm has a Laplace distribution for the particles' positions. The special case of harmonic oscillator, where a parabolic potential distribution was assumed, led to the interesting Gaussian distribution of the harmonic oscillator in Table I. It has been reported in literature [22] that the classical PSO algorithm exhibits a Gaussian behavior, which can be used to simplify the canonical form of the method by using position-only update based on the Gaussian distribution. Thus, it seems that the quantum method can introduce a derivation for the classical case if the "correct" potential distribution, the parabolic law in equation (14), is chosen. The interpretation of the optimization algorithm that is based on quantum mechanics could open the door for a plenty of possible probability distributions with various physical insight for each of these distributions.

## VIII. Application of the QPSO Algorithm to Array Antenna Synthesis Problems

Both of the classical PSO and the QPSO were tested first against a set of benchmark functions. The functions tried include the sphere function, Rastrigin's function, and Rosenbrock's valley [3], [5]. Other functions were used to examine specific aspects in the performance of the algorithm. Asymmetric initialization ranges were employed to test the origin-bias seeking of the optimization. The algorithm performance in these problems was satisfactory enough to start applying the method to practical problems.

In improving the classical PSO to test its competence with the new QPSO scheme, the same level of complexity in both algorithms is maintained. That is, although there are some advanced supervisory methods suggested in literature to enhance the convergence of the PSO under varying possible physical problems [2], [13]-[14], such methods increase the complexity – and thus the computational cost – of the algorithm. Therefore, our comparison is based on maintaining the same level of simplicity in both schemes to guarantee reasonable measures for the performance of the new algorithm compared with the classical version. Also, several runs with varying seeds for the random number generators were reported in the comparison.

### A. General Description of the Linear Array Synthesis Problem

We consider now the application of the QPSO algorithm to linear array antenna design problems. In this paper, the consideration focuses on shaping the main beam and the side lobe pattern of the array to meet desired characteristics given by a user-defined function $T(u)$. Suppose that we have $N$-element linear array separated by a uniform distance $d$ (taken in this paper to be $d = \lambda/2$). The normalized array factor is given by

$$AF(u) = \frac{1}{AF_{\max}} \sum_{n=1}^{N} I_n e^{j 2\pi n d u / \lambda} \quad (25)$$

where $I_n$ are the amplitude coefficients (generally complex), and $u = \sin\theta$, where $\theta$ is the angle from the normal to the array axis. *AF* stands for the array factor and $AF_{\max}$ is the maximum value of the magnitude of the array factor defined as $AF_{\max} = \max\left\{\left|AF(u)\right|\right\}$, where the maximization operator is carried over the observational space *u*.

Fig. 3 shows a block diagram representation of the linear array antenna synthesis problem. The algorithm, whether classical or quantum, works on particles moving in *N*-dimensional hyperspace. The information about amplitudes and phases are encoded in the coordinates of the particle that are normalized between 0 and 1. The mapping block shown in Fig. 3 gives the required transformation between the abstract optimizer and the real physical problem.

We will compare the performance of quantum and classical PSO algorithms in achieving certain desired patterns. To establish the comparison on a fair ground, a *clipping criterion* (hard domain) is employed in both algorithms that well not allow candidate solutions to pass certain boundaries [12]. If a particle hits a specific boundary, its value will be fixed at the value of that boundary, and eventually it should go back to the solution space searching for the global minimum that exists there. All comparisons between different algorithms were reported using the same number of iterations and population size. Preliminary experiments with the QPSO algorithm, when applied to benchmark functions, suggest using high number of population size in order to achieve a satisfactory performance. Based on this we fix the number of populations to 60 particles in the coming examples. This enhances the "averaging" or the smoothing of the convergence curves obtained. In the following examples, various realizations of the same experiment – starting with different seeds - produced results that are close to each other. An averaging of ten runs is considered and found to be sufficient.

The number of fitness calls equals $N_{\text{iteration}} \times N_{\text{population}}$, where $N_{\text{iteration}}$ is the number of iterations and $N_{\text{population}}$ is the number of population. Thus, increasing the population size will dramatically increase the number of fitness calls. Since the calculation of the objective function is the bottleneck in most EM applications, this will make the optimization process increasingly expensive. Therefore, although increasing the number of particles in the swarm will intuitively increase the possibility to reach the global optimum, for a practical EM problems the population size should be kept as small as possible.

### B. Optimizing Sidelobe Patterns

The objective function used here is not directly proportional to the absolute difference between the obtained and the desired pattern. Instead, the "Don't exceed criterion" has been utilized in the formulation of the objective function. That is, an error will be reported only if the obtained array factor exceeds $T(u)$, the desired side-lobe level. We start by generating $M$ observation points in the sine space $u = \sin\theta$. Then, at each point we compare between the obtained array factor $AF$ and $T$. If $\left|AF\left(u_k\right)\right|$ exceeds $T\left(u\right)$, then we store the error $e\left(u_k\right) = \left|AF\left(u_k\right)\right| - T\left(u_k\right)$. We end up with a vector of errors $\mathbf{e}$ that has nonzero positive values only at the locations where the obtained array factor exceeds the desired level. The cost function $F$ is defined then as the averaged sum of the squares of the errors obtained and is expressed as

$$F = \frac{1}{M}\sum_{k=1}^{M}\left[\max\left(\left|AF\left(u_k\right)\right| - T\left(u_k\right), 0\right)\right]^2 \tag{26}$$

where $M$ indicates the number of sampling points and $\Delta u = 2/\left(M-1\right)$.

Fig. 4 illustrates the optimization of 40-element linear array antenna with tapered side-lobe performance. In Fig. 4(a) the obtained array pattern is shown together with the resulted current distribution. In this problem we exploited the symmetry of the current distribution since the desired envelope is already symmetric. This reduces the degrees of freedom into half of the number of array elements.

Both of the classical PSO algorithms, used for the comparison, and the proposed QPSO, employ a*synchronous* velocity update. That is, the update of the global best performance is done after each particle's move, rather than waiting until the required information is collected from the entire particle swarm at the end of the current generation [10]. This technique has been adopted in electromagnetic applications [3]-[4] but very recently some researchers [11] reported a detailed study showing that it is more powerful than synchronous update most of the time. Moreover, in the classical PSO version employed here we integrated what we defined as *hard* domain instead of *soft* domain, leading to an improved classical PSO version [12]. In this version the value of the maximum velocity $V_{\max}$ is changed to enhance the performance. Also, the reflection boundary condition (RBC) is combined with the hard domain. Fig. 4(b) shows clearly that when PSO is run without RBC, the algorithm found itself trapped in a local minimum while the QDPSO achieves much better solutions. Thus, by enabling the RBC feature and changing $V_{\max}$ we could obtain much better solution; however, the best solution at $V_{\max}$ = 0.15 could not compete with the QDPSO, which was obtained using simpler tuning strategy. The only parameter $g$ of the algorithm was tuned by direct trial and error strategy through testing problems with small number of generations. Then, we run the optimization, using the best $g$ obtained, for high number of iterations. This is much easier compared with the genetic algorithms (GA) and classical PSO where several tuning parameters should be chosen, each with various possible values and directions of change.

Fig. 5 shows the comparison between PSO and QPSO for more challenging synthesis requirement. Here we impose a wide null in the side-lobe region. The performance of the QDPSO and the classical PSO is reported. The quantum version scores much better for the same number of iterations and population size. In Fig. 5(b) we also show the effect of varying the characteristic length $g$ linearly from 2.5 to 4. It is clear that the QDPSO achieves better convergence rate than the PSO.

The linear variations of $g$ can be justified conceptually if one notices that by such variation the corresponding characteristic length of the well will decrease with iterations. As we discussed before, by decreasing the characteristic length, we increase the probability of finding the particle around the origin, thus highlighting the *local* search capabilities of the algorithm. This enhances the performance since with increasing iterations the particles become more close to the global minimum. If one is interested in conducting the comparison only through the first 1200 iterations, then it seems then that by allowing linear variation of the control parameter the algorithm looks by itself for the best $g$. The user can later try different values for the minimum and maximum limits of this parameter to narrow the search range, provided that improvement in the performance is possible.

### C. Synthesized Main Beam Patterns

Shaping main beams requires minimizing the *absolute* difference between the obtained array pattern and the desired pattern shape. The cost function is defined as

$$F = \frac{1}{M}\sum_{k=1}^{M}\left[\left|AF\left(u_k\right)\right| - T\left(u_k\right)\right]^2 \tag{27}$$

Using Don't Exceed criterion in the side lobe region and the above absolute error criterion in the main beam region leads to multi-objective optimization. In general, this will increase the difficulty of the problem, reducing the chances of reaching to the global minimum. The authors have tried to use the aggregation method by transforming the two cost functions into the following single cost measure $F = \alpha_1 F_1 + \alpha_2 F_2$, were $F_1$ and $F_2$ are the cost functions in the main beam and the side lobe regions, respectively. The constants $\alpha_1$ and $\alpha_2$ are the possible weights of each objective function, $F_1$ and $F_2$, respectively. Optimizing the problem this way did not lead to satisfactory results with either the quantum or the classical PSO algorithms. Some prior knowledge about the weights above is needed to formulate the cost function. It is clear that those multi-objective optimization problems cannot lead in general to solutions that satisfy all sub-measures simultaneously. To solve this problem, a single objective function, as defined by (27), is used in the entire region. The results are shown in Fig. 6 where the goal is to synthesize *asymmetrical* pattern by using both the amplitude and phase of each array element. Notice that since we are trying to force the array pattern to follow the desired target, there will be certain amount of ripples around this target. This is in contrast to the previous case when the Don't Exceed criterion has been used. From Fig. 6(b) we can see that the QDPSO is much faster in convergence compared with the classical version.

In our experiments, the QDPSO appears to be more stable and easy to tune compared with the quantum harmonic oscillator version (QOPSO). The performance of the square well (QSPSO) algorithm was poor for most of the cases tried. The authors noticed that by selecting the width of the square well to be very small, the potential distribution approaches the delta function and the performance improves. However, according the *pdf* of the QSPSO in Table I, the corresponding density function shrinks also by the same ratio, thus increasing the probability of finding more particles around the origin. This leads to very fast convergence but with higher chances to be trapped in local minima since fewer particles are allowed to "spy" on other regions in the solution domain. In some optimization problems one may be interested in very fast convergence rather than achieving global low cost function (global solution). Thus, the QSPSO can be utilized in such applications.

In spite of the above difficulties with the QOPSO and the QSPSO, It is noticed that these algorithms might compete in other types of applications. Thus, it is up to the user to make the choice of the proper version of QPSO to better suite his problem. Such results are omitted from this paper for brevity.

## IX. Infinitesimal Dipoles Equivalent to Practical Antennas

### A. Formulation of the Problem

To demonstrate the applicability of the new quantum PSO method to different types of applications, we consider in this section the problem of finding a set of infinitesimal dipoles that can model an arbitrary practical antenna configuration, which are realistic antennas with practical gain and radiation patterns. The modeling here means that the obtained set of dipoles must be capable of producing both the near and far fields of the antenna under consideration. This concept was introduced in [15] with both electric and magnetic dipoles used to model the near field of a radiating structure. In [16], the problem was formulated as an optimization problem solved using evolutionary genetic algorithm (GA) techniques. The GA was introduced to tackle a general inverse source modeling problem using infinitesimal dipoles with both electric and magnetic types [17]. However, the main purpose there was to predict the far field performance of a device under test (DUT) based on the measured near field. In this section, starting from a known near field distribution of an antenna, infinitesimal dipoles are obtained to provide the same near field and consequently the same far field.

Assume that the antenna has the near field distribution $\left(\mathbf{E}^a, \mathbf{H}^a\right)$ at a certain surface $\boldsymbol{S}$. Assume also a set of dipoles $\{\chi_i\}_{i=1}^{N}$, where $N$ is the number of dipoles and $\chi_i$ is a 9-element vector representing the parameters of the $i$th dipole given by

$$\chi_i = [\mathrm{Re}\{M_u^i\},\quad \mathrm{Im}\{M_u^i\},\quad r^i]^T \qquad u = x, y, z \quad \text{and} \quad r^i = (x^i, y^i, z^i) \tag{28}$$

Here $M_x^i$, $M_y^i$, and $M_z^i$ are the $i$th dipole moments located at the position $x^i$, $y^i$, and $z^i$, which are constrained by the actual antenna size. The fields generated by the dipoles are denoted by $\left(\mathbf{E}^d, \mathbf{H}^d\right)$. The cost function is defined as follow

$$F = \left[ \frac{1}{\sum_{n=1}^{N_{ops}} \sum_{u} \left| E_u^a(\mathbf{r}_n) \right|^2} \sum_{n=1}^{N_{ops}} \sum_{u} \left| E_u^a(\mathbf{r}_n) - E_x^d(\mathbf{r}_n) \right|^2 \right]^{1/2} \tag{29}$$

where $N_{ops}$ is the number of observation (samples) points, and $\mathbf{r}_n$ is the position vector of the *n*th sampling point where these set of points are assumed to sample the observation surface ***S***. Notice that the objective function defined in (29) uses only the electric field. Once the electric field is synthesized by the obtained set of infinitesimal dipoles, the magnetic field will be satisfied automatically because both fields must satisfy Maxwell's equations.

In general, this cost measure is highly nonlinear, with an objective function landscape full of local minima, making the optimization problem very difficult unless a powerful global search method is used. Once the infinitesimal dipoles are found, both the near and far fields can easily be computed.

**B. Infinitesimal Dipole Model of Circular Dielectric Resonator Antenna**

The above procedure is applied to a dielectric resonator antenna (DRA). Fig. 7(a) shows the configuration of a circular DRA located above an infinite ground plane. A coaxial probe excites the antenna as shown in Fig. 7(b). An accurate Method of Moment (MOM) procedure [18] is used to analyze the structure. The DRA is tuned to resonate at the frequency 10 GHz unlike the sample used with the GA [19], which was off-resonance. Near-field data are calculated at a square plane of side length 1.0 $\lambda$. The distance of the observation plane from the ground is taken to be 1.0 $\lambda$ as shown in Fig. 8(b). A set of 10 dipoles is considered. The QDPSO algorithm is used to minimize the objective function defined in (29) with the control parameter $g$ set to 3.0 and population size of 80 particles. The dipoles locations are restricted to be inside the physical domain of the DRA while the dipole moments are chosen based on the order of magnitude of actual near field data. Fig. 8 shows the convergence curve obtained for 5000 iterations. Samples of the obtained near fields are compared with the actual near fields as shown in Fig. 9(a). Excellent agreement is observed. The RMS error (defined in [19]) computed for this problem is less than 3%. Fig. 9(b) shows the comparison between the far-field radiation patterns of the equivalent dipoles and the actual antenna. It should be mentioned that the examples in [19] used only 5 dipoles and the RMS errors was above 7%.

In [17] it was reported that eight dipoles constitute a practical upper limit beyond which convergence problems will start to affect the GA optimization process. However, the new quantum method was able to predict a solution for 10 dipoles using only one control parameter, which doesn't need tuning. The value of the control parameter $g = 3.0$ was used directly based on the previous experience with benchmark functions and the array antenna synthesis problems.

**C. Circuit Model of Circular Dielectric Resonator Antenna**

The antenna shown in Fig. 7 represents a resonator coupled with the source through a feeding mechanism. Such an antenna can be modeled by lumped-element *RLC* resonant circuits that resonate at the same frequency of the actual antenna [20]-[21]. The probe coupling is considered in the circuit model by adding the series $L_c$ and $C_c$ as shown in Fig. 10. The problem is formulated to minimize the square difference between the input impedance of the equivalent circuit, $Z_{in}^{e}$, and the input impedance of the actual antenna, $Z_{in}^{a}$, using the following objective function

$$F = \left[ \frac{1}{\sum_{n=1}^{N} |Z_{in}^{a}(f_n)|^2} \sum_{n=1}^{N} |Z_{in}^{e}(f_n) - Z_{in}^{a}(f_n)|^2 \right]^{1/2} \tag{30}$$

where $N$ is the number of frequency points and $f_n$ is the $n$th frequency sample. This problem requires optimizing five variables, $R$, $L$, $C$, $L_c$ and $C_c$. The ranges of these variables must be carefully chosen since they are not independent. For example, for the parallel resonant circuit, the resonance frequency is given by $f_0 = 1/\sqrt{LC}$ . Since the actual impedance of the DRA shows resonance at 10 GHz, then both $L$ and $C$ are related to each other. It is assumed that only one mode is excited, which is equivalent to the parallel *RLC* resonator. Therefore, we considered a small frequency band around the resonance frequency of the antenna. The QDPSO was used to solve the problem with 30 particles, $g$ = 3.0, and 4000 iterations. The range of the parameters and the final values obtained by the optimization are shown in Table II.

**Table II** The parameters range for each circuit element and the final values obtained after finishing the optimization.

| Element\Min, Max, Obtained | Minimum Value | Maximum Value | Obtained Value |
|---|---|---|---|
| $R$ ($\Omega$) | 1.00 | 100.0 | 56.590 |
| $L$ (nH) | 0.01 | 1.0 | 0.0724 |
| $C$ (pF) | 1.00 | 10.0 | 3.4864 |
| $L_c$ (nH) | 0.10 | 10.0 | 0.5601 |
| $C_c$ (pF) | 0.10 | 10.0 | 0.3651 |

Fig. 11(a) illustrates the convergence curve QDPSO algorithm with 30 particles, $g$ = 3.0, and 4000 iterations. The comparison between the input impedance obtained using MOM [18] and the one obtained using the equivalent circuit of Fig. 10 is presented in Fig. 11(b). It is clear that the predicted input impedance is in excellent agreement with the accurate MOM calculation. Furthermore, by processing the obtained equivalent circuit, it is possible to calculate the unloaded resonance frequency and the radiation quality factor as 10.0164 GHz and 12.4161, respectively. The coupling coefficient and the loaded quality factor are also calculated to be 1.1183 and 5.8613, respectively [21].

## X. Conclusion

A generalized framework for Quantum Particle Swarm Optimization (QPSO) suitable for electromagnetic problems was proposed. Through choosing appropriate attractive potential fields, different algorithms were obtained by solving the corresponding Schrödinger equations. The physics of the quantum swarm helped in providing a theoretical explanation of why the algorithm works.

Considering the number of iterations and the population size as common requirements in every evolutionary computing algorithm, we found that the QPSO algorithm contains only one control parameter while its classical counterpart requires the use of four parameters ($c_1$, $c_2$, $w$, and $V_{\max}$) for satisfactory convergence to the desired solution. Although boundary conditions were not considered before as a main factor in determining the performance of the PSO, this paper suggested that a suitable choice of this condition could be critical in some applications. The computational cost of tuning the four numerical parameters in the classical PSO were very high compared with single-parameter QPSO. In addition, by linearly increasing $g$, faster convergence could be obtained, but not necessary to global optimum, which could be useful for certain applications when one is looking for sub-optimal solutions. Therefore, the tuning of the QPSO is much simpler and more attractive compared to many of the classical versions of PSO.

The applicability of the new optimization procedure was investigated by studying the problem of synthesis of linear array antennas and the performance of the QPSO was shown to outperform the classical PSO most of the time in the convergence rate as well as the final error level. The QDPSO method was utilized to simulate an actual antenna, such as the circular DRA, by a set of ten infinitesimal dipoles that provide similar near and far fields, which outperformed previous solutions using the GA method applied to the same problem. Also, the QDPSO was utilized to find a circuit model for the DRA. The resonant circuit was used to predict several parameters such as the resonant frequency, the quality factors, and the coupling coefficient.

$$
\begin{array}{l}
\text{Inialize } x^{m},\ \mathbf{P}^{m}_{\text{local}},\ \mathbf{P}_{\text{global}} \\
\text{Do i} = 1, N_{itr} \\
\quad \text{Do } m = 1, N_{\text{population}} \\
\qquad \text{Update } \mathbf{P}^{m}_{\text{local}} \text{ and } \mathbf{P}_{\text{global}} \\
\qquad \varphi_1 = rand(0,1),\ \varphi_2 = rand(0,1) \\
\qquad \mathbf{P} = \dfrac{\varphi_1 \mathbf{P}^{m}_{\text{local}} + \varphi_2 \mathbf{P}_{\text{global}}}{\varphi_1 + \varphi_2} \\
\qquad u = rand(0,1) \\
\qquad L = L(g, u, |x^{m} - p|) \\
\qquad \text{if } rand(0,1) > 0.5 \\
\qquad\quad x = p + f(L, u) \\
\qquad \text{else} \\
\qquad\quad x = p + f(L, -u) \\
\qquad \text{end if} \\
\quad \text{end Do} \\
\text{end Do}
\end{array}
$$

**Fig. 1** Pseudo-code for the QPSO algorithm with the explicit form of the functions $L$ and $f$ the QDPSO, QOPSO, and QSPSO as described by Table I.

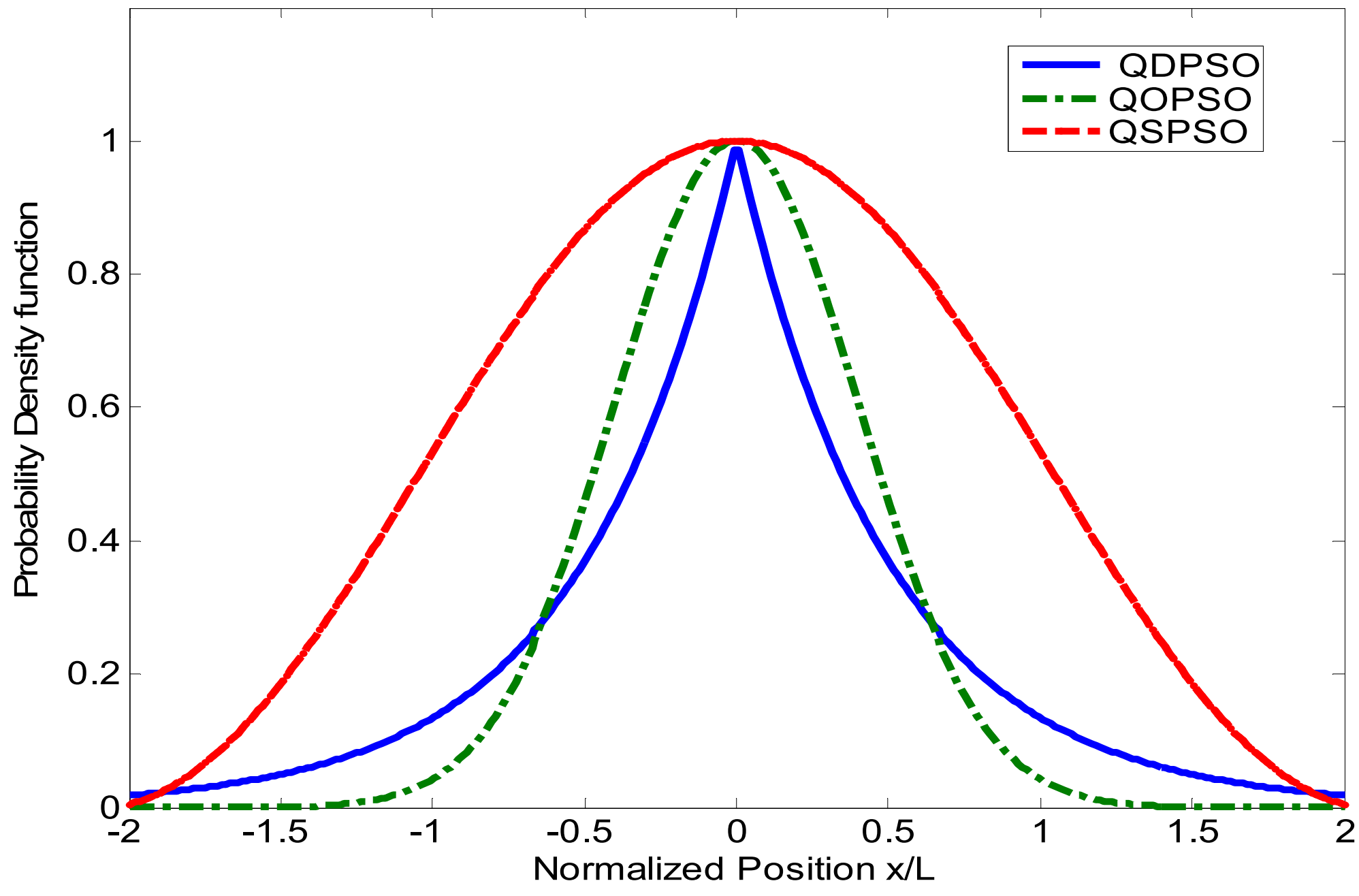

**Fig. 2.** Comparisons between the probability density functions of the three potential wells introduced in this article (Delta, Harmonic and Square). The position is normalized to the characteristic length of each potential well.

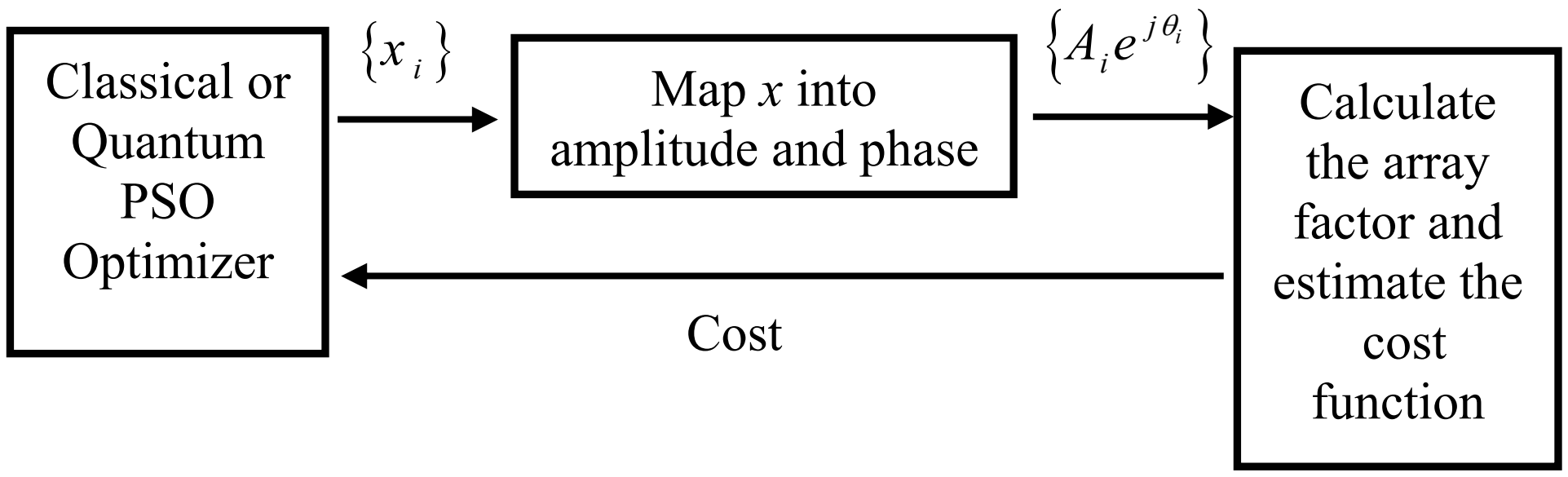

**Fig. 3.** Block diagram representation for the problem of linear array synthesis. Here the particles positions are $x_i \in \{0 \le x_i \le 1, i = 1,2,...,N\}$. Amplitudes and phases are ranged as $A_i \in \{A_{\min} \le A_i \le A_{\max}, i = 1,2,...,N/2\}$ and $\theta_i \in \{0 \le \theta_i \le 2\pi, i = 1,2,...,N/2\}$, respectively.

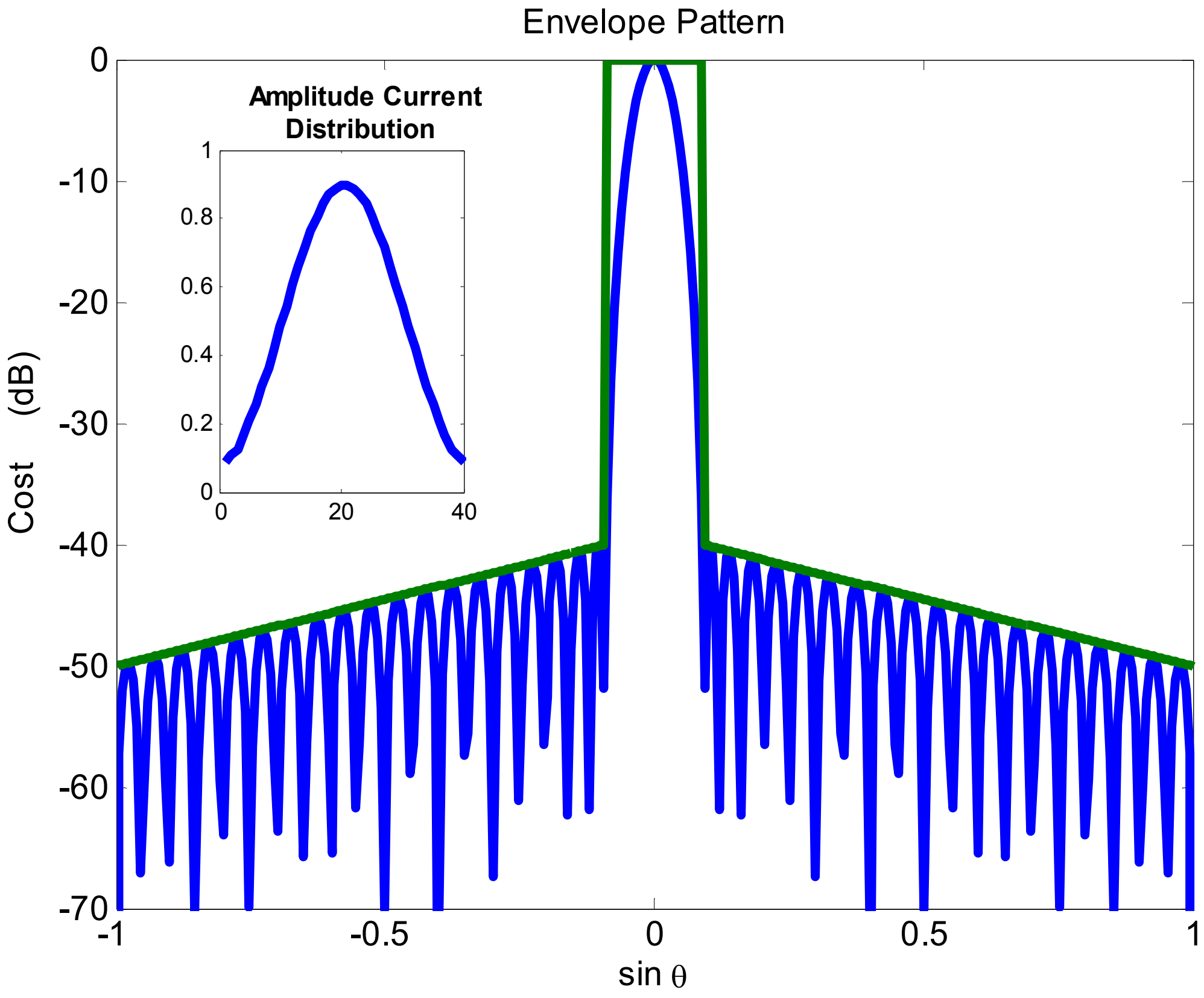

(a)

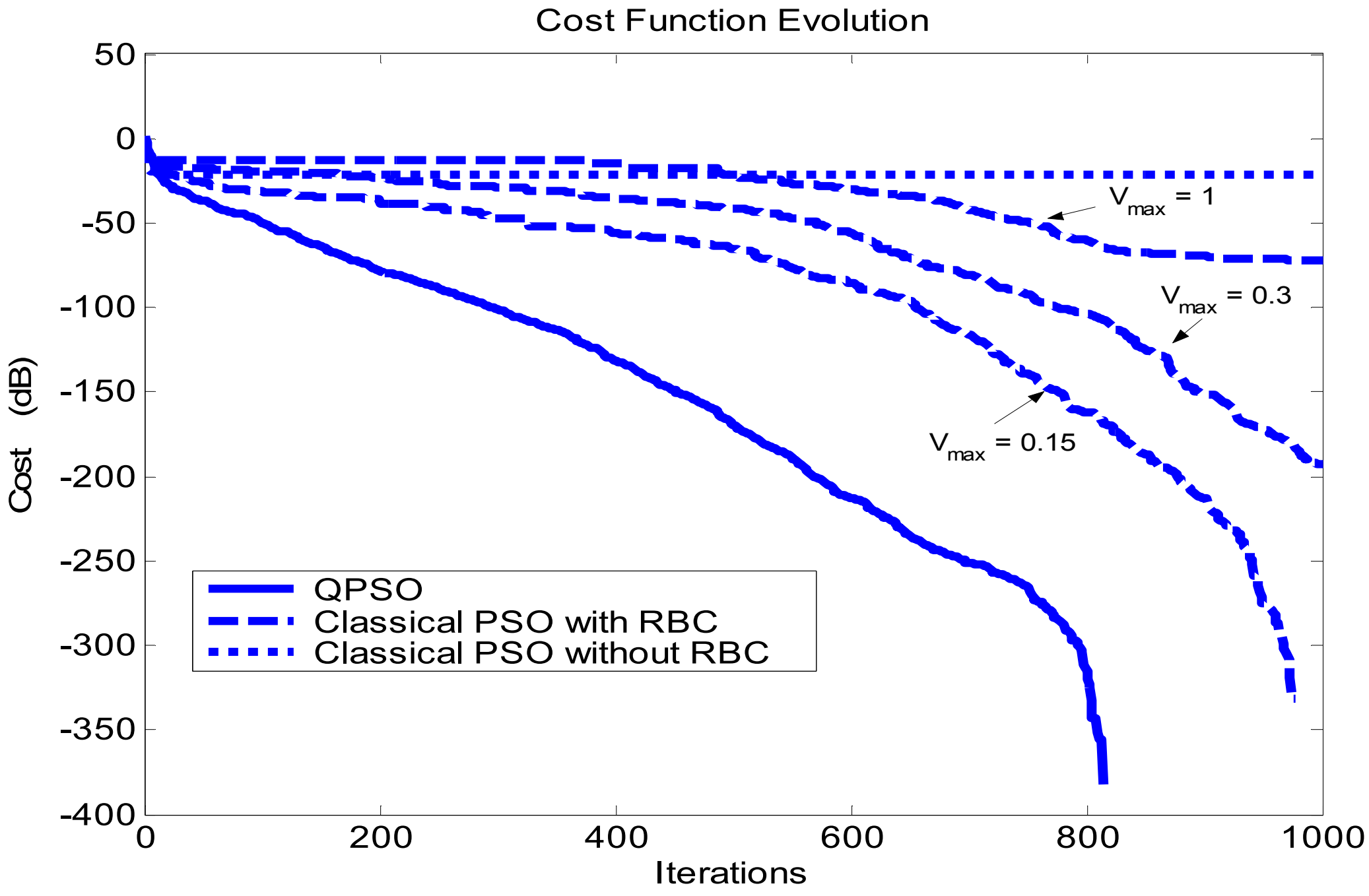

(b)

**Fig. 4** Amplitude-only synthesis with tapered side lobes and enforced symmetry. 40 element and beamwidth of $9.5^o$ (a) The obtained pattern and the current distribution using QDPSO algorithm with 1000 iterations, population size of 60, 301 observation points, and characteristic length of $g$ = 3 (b) Convergence curves for the QDPSO and the classical PSO. The classical PSO uses the same number of iterations and population size. The inertia weight $w$ is varied linearly from 0.9 to 0.2, $c_1 = c_2 = 2.0$, and the maxim velocity limit $V_{\max}$ is changed as indicated.

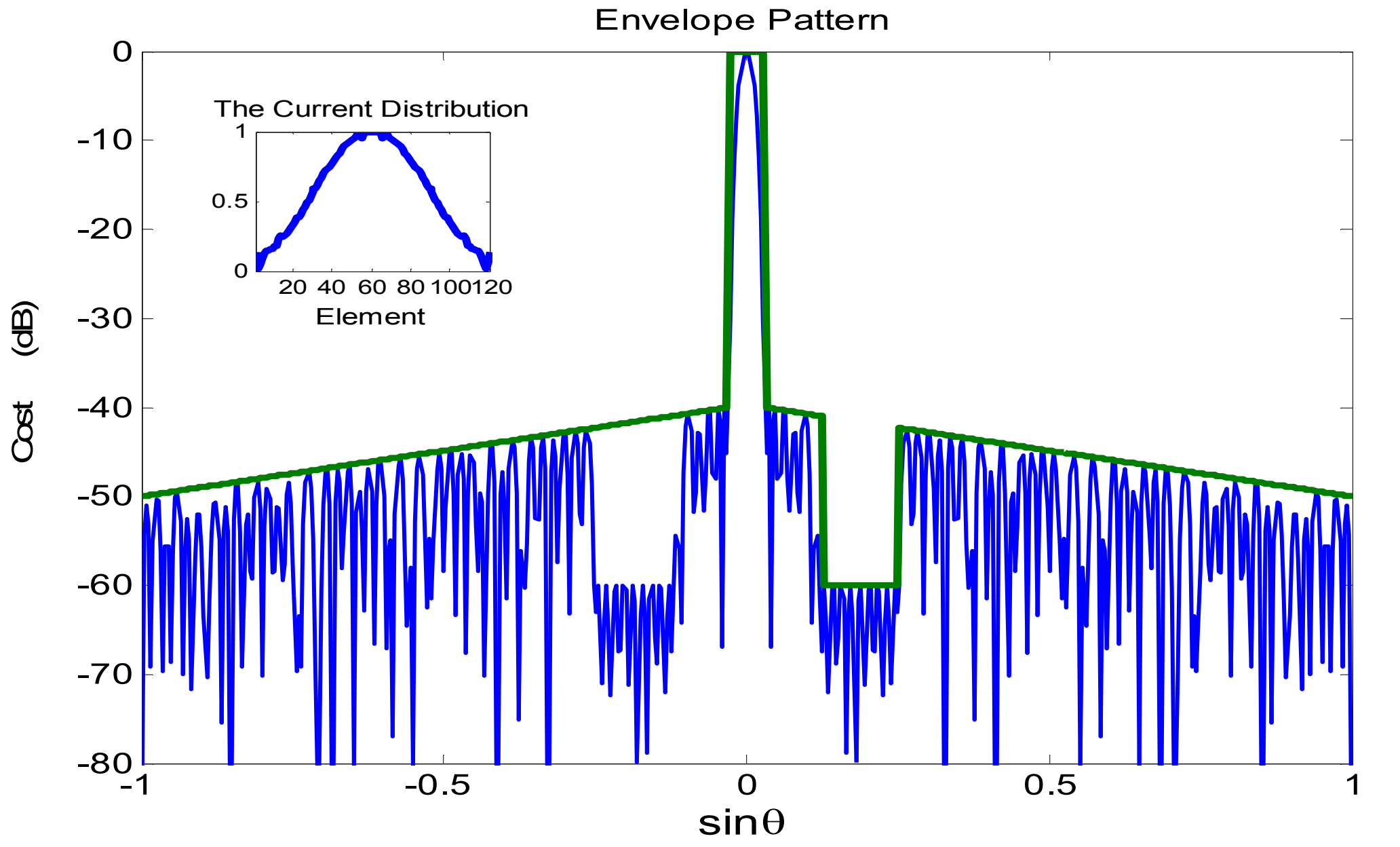

(a)

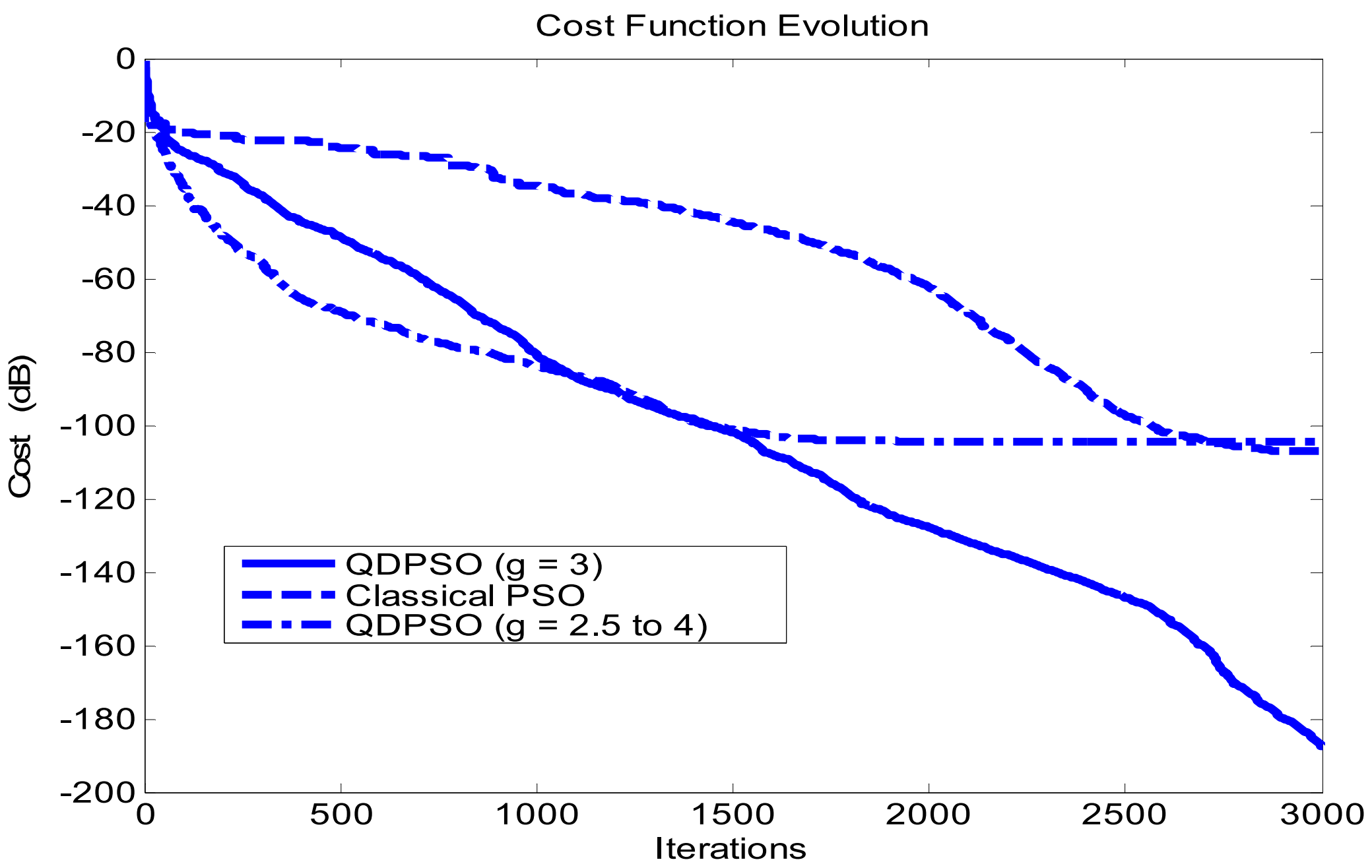

(b)

**Fig. 5**. Amplitude-only (enforced symmetry) synthesis with tapered side-lobe and deep null. 120 elements with $3^o$ beamwidth (a) The obtained pattern and the current distribution using QDPSO algorithm with 3000 iterations, population size of 60, 401 observation points, and characteristic length of $g$ = 3 (b) Convergence curves for the QDPSO and the classical PSO. The classical algorithm uses the same number of iterations and population size. The inertia weight $w$ is varied linearly from 0.9 to 0.2, $c_1 = c_2 = 2.0$, and $V_{max} = 0.3$ with RBC.

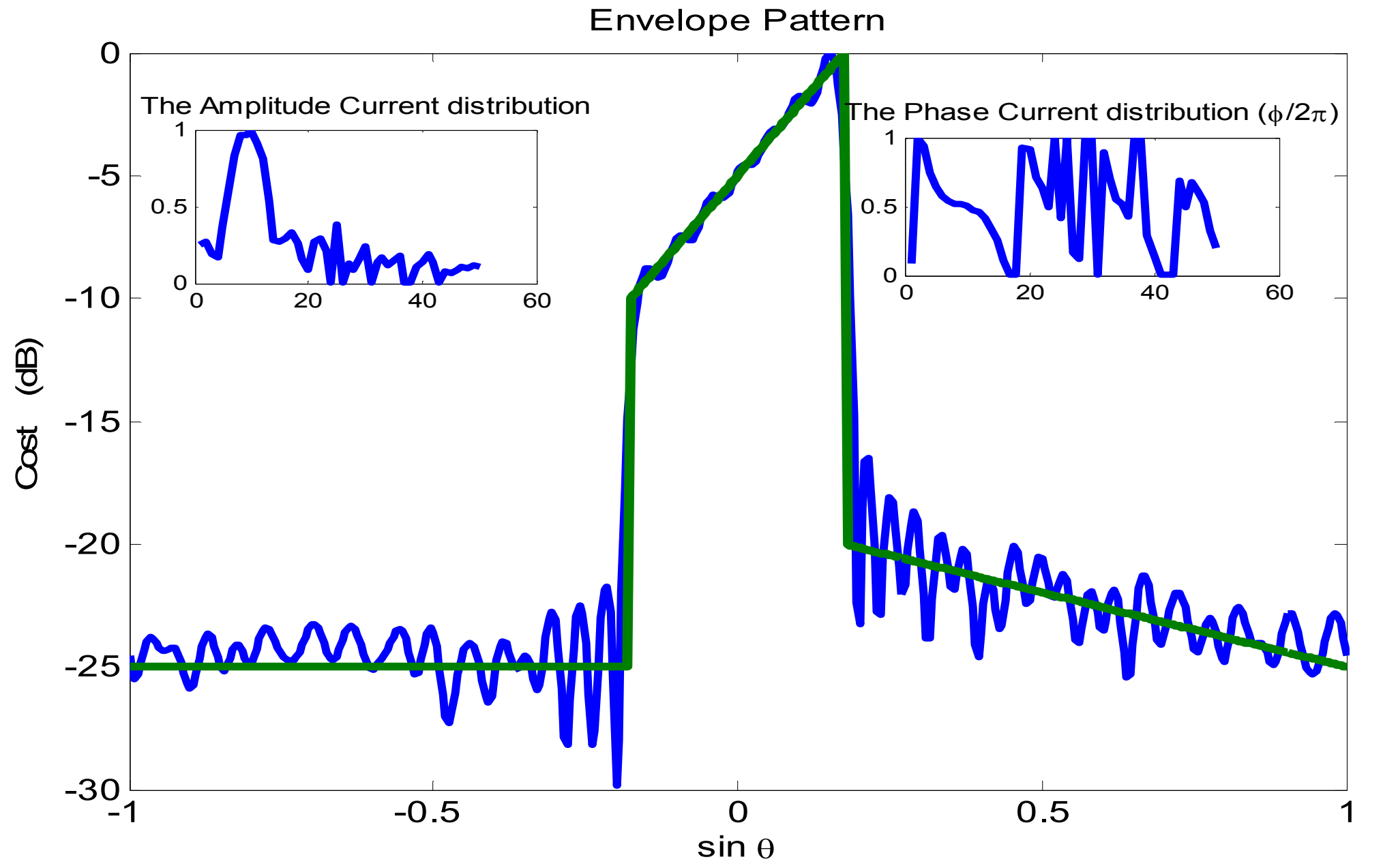

(a)

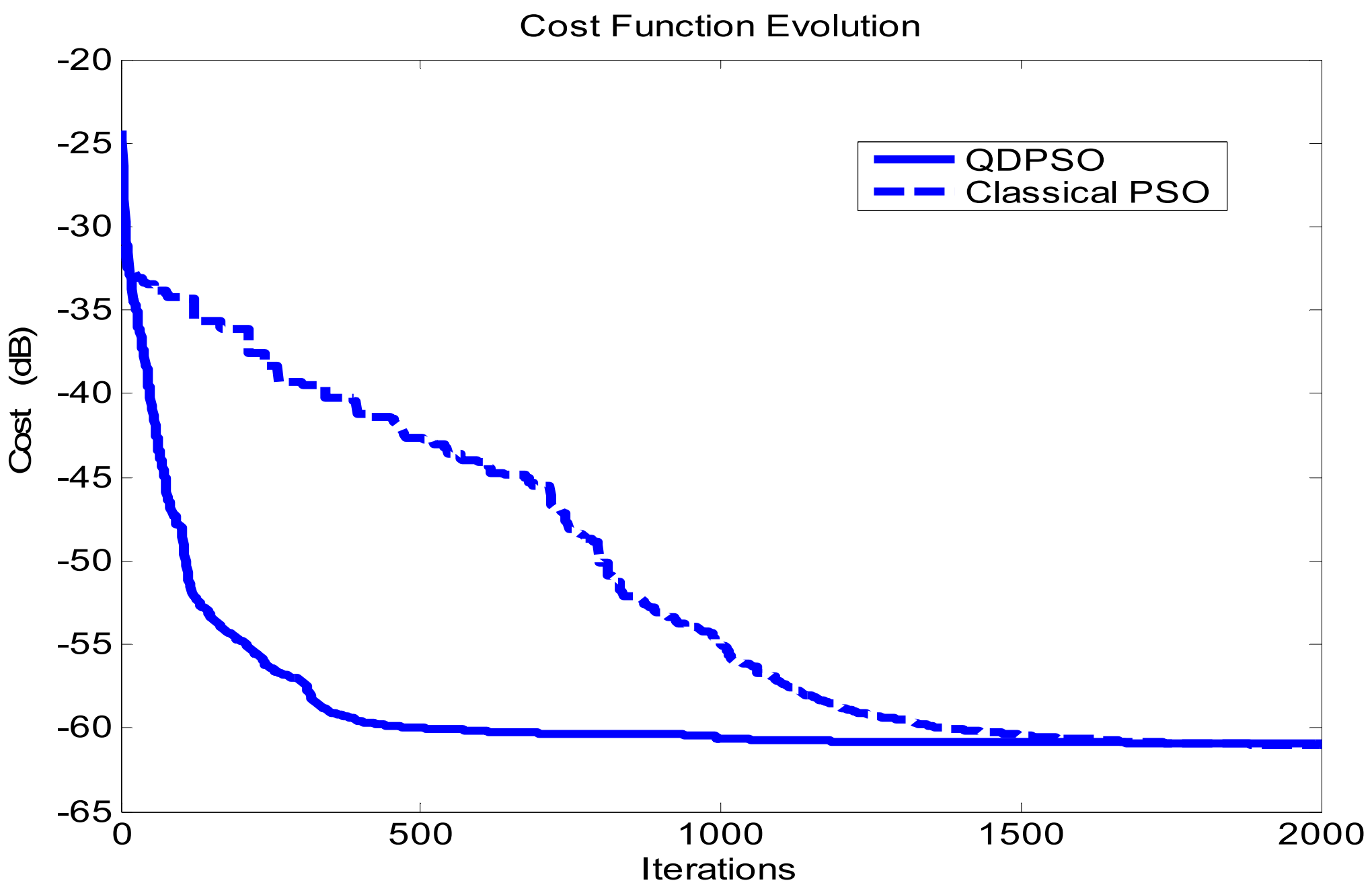

(b)

**Fig. 6** Complex synthesis, 50 element with beamwidth of $20^o$ (a) The obtained pattern and the current distribution using QDPSO algorithm with 2000 iterations, population size of 60, 501 observation points, and characteristic length of $g = 3$ (b) Convergence curves for the QDPSO and the classical PSO. The classical algorithm uses the same number of iterations and population size. The inertia weight $w$ is varied linearly from 0.9 to 0.2, $c_1 = c_2 = 2.0$, and $V_{\max} = 0.3$ with RBC.

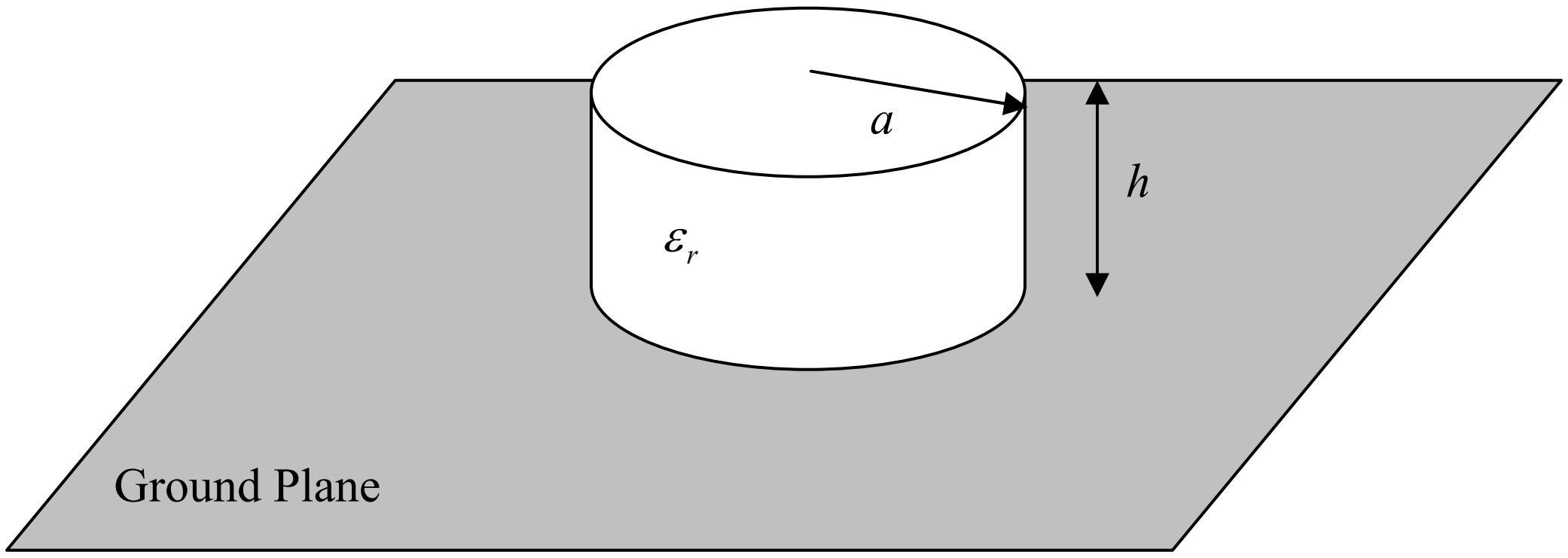

(a)

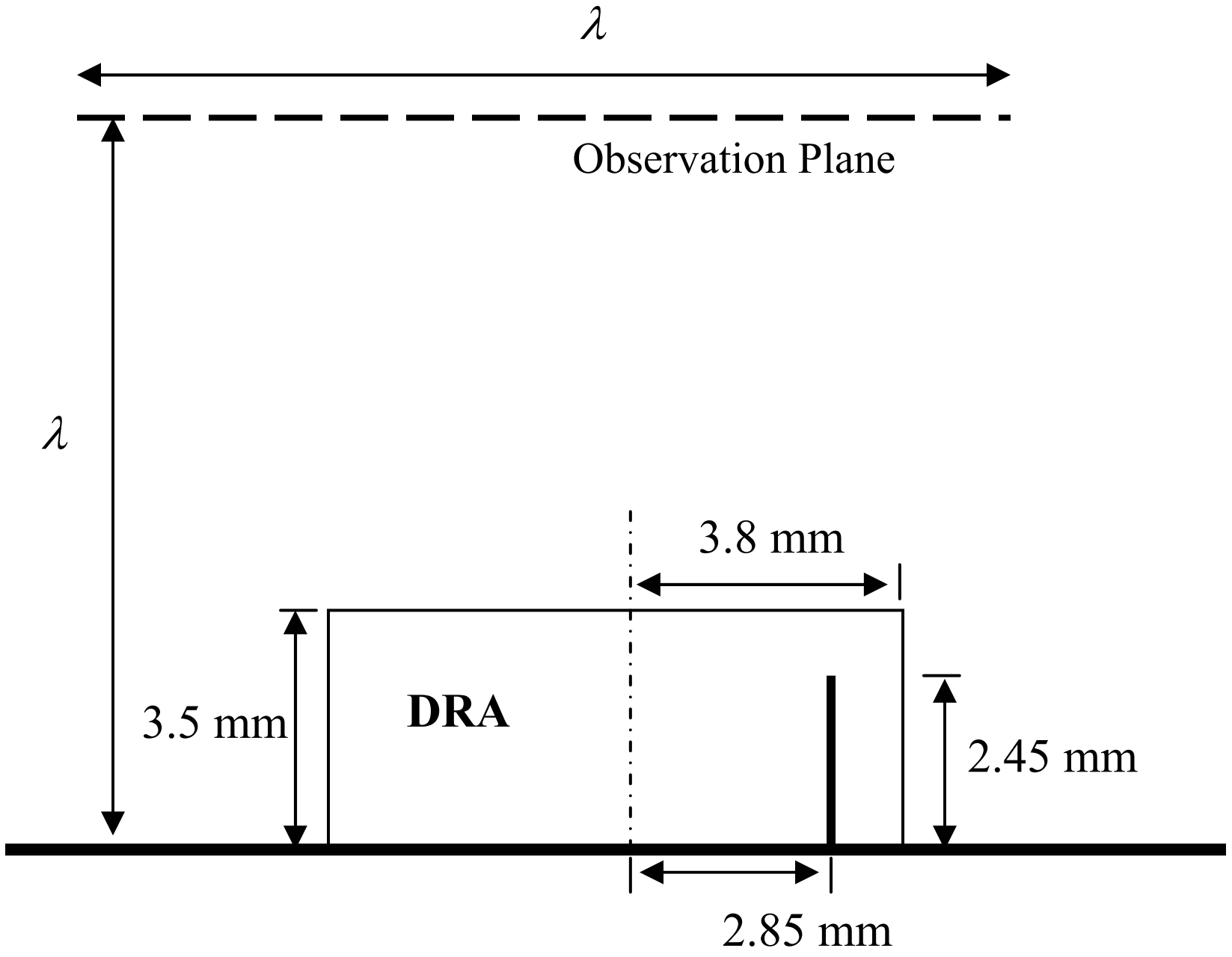

(b)

**Fig. 7** (a) Geometry of a circular dielectric resonator antenna located above an infinite ground plane. (b) Cross sectional view illustrating the dimensions of the DRA and the location of the monopole deep within the DRA. The dielectric constant of the DRA material is $\varepsilon_r = 10.2$.

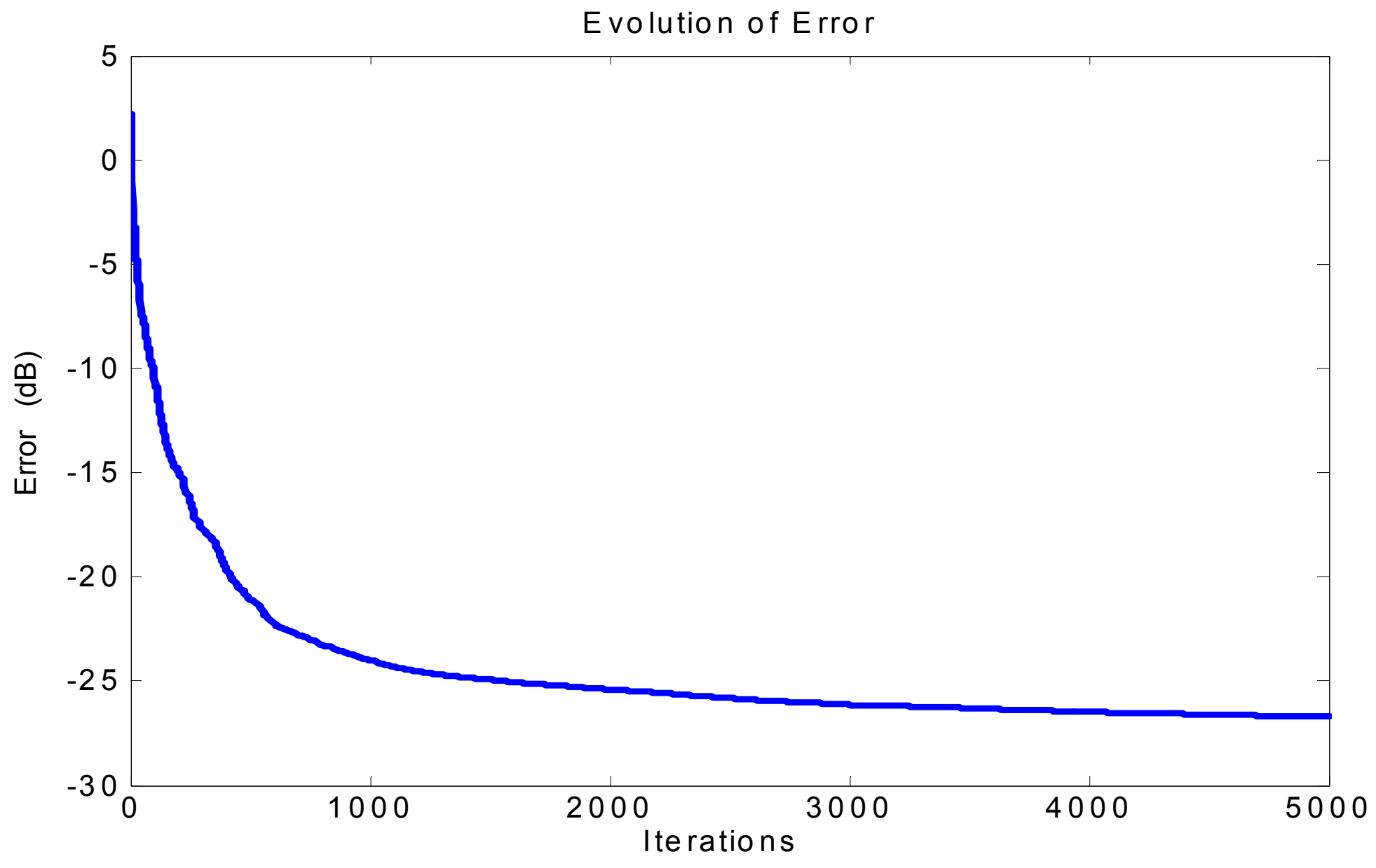

**Fig. 8** Convergence performance of the DRA synthesis problems using a set of 10 dipoles with population size of 80 particles and $g = 3.0$.

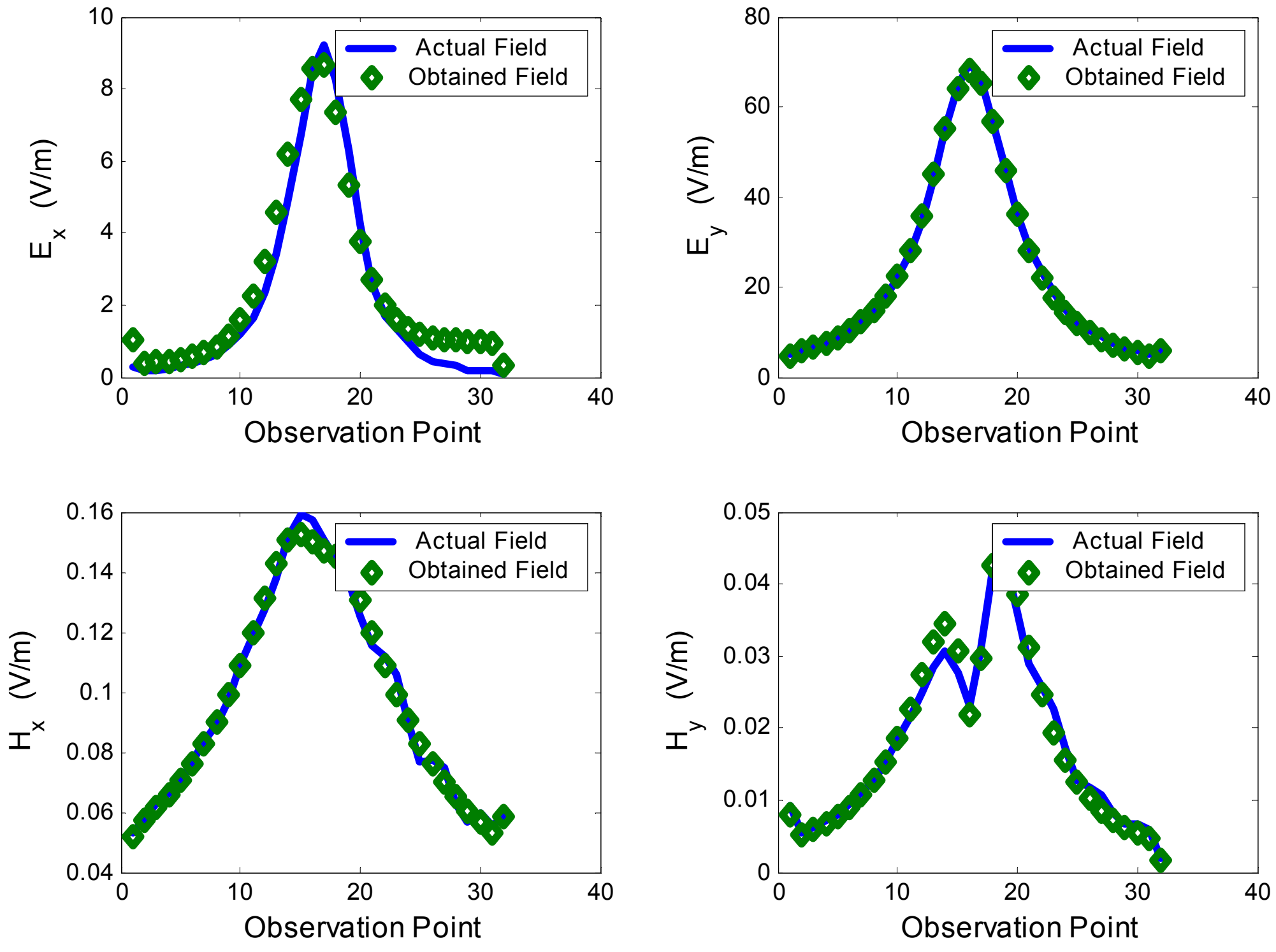

(a)

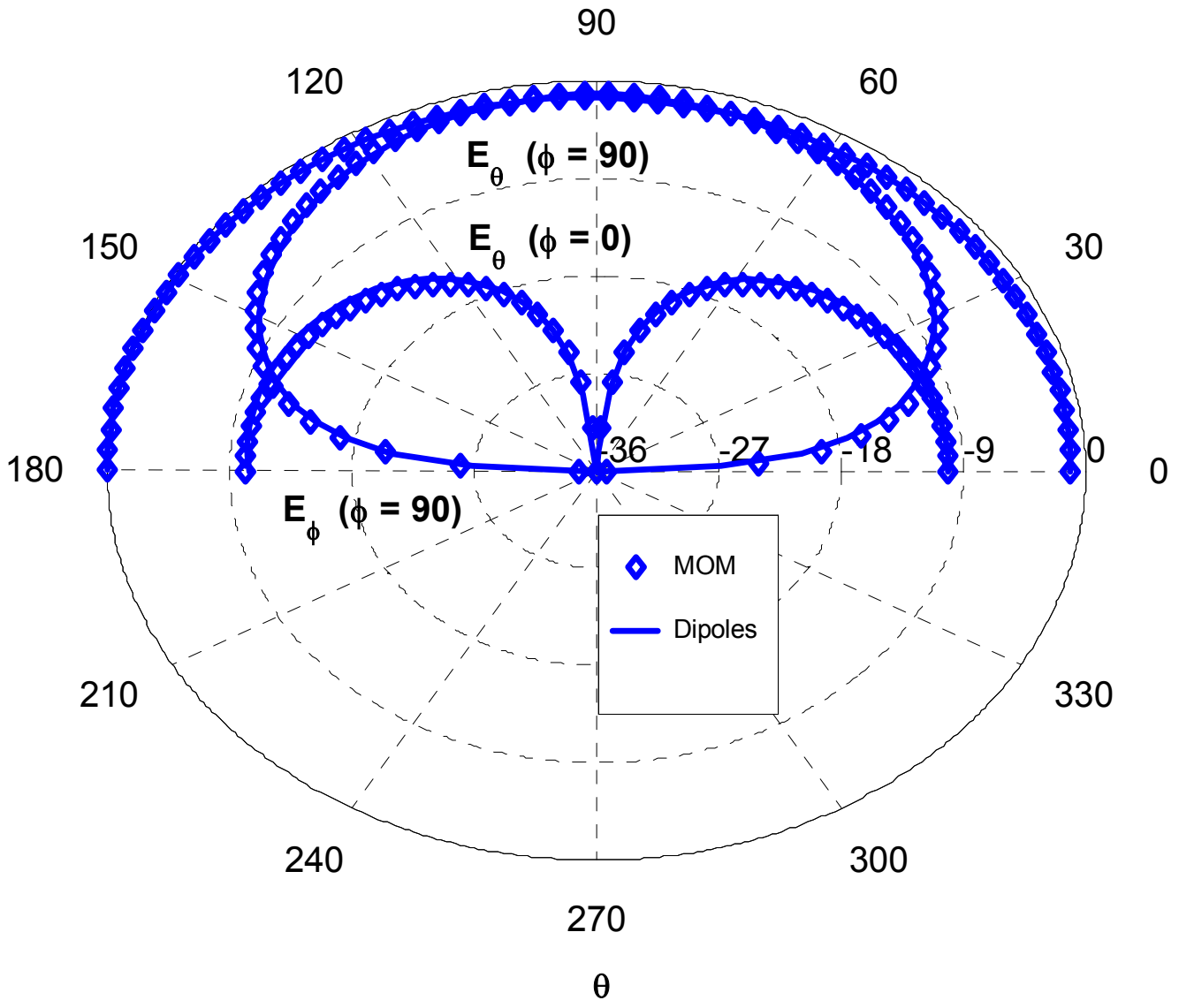

(b)

**Fig. 9** Comparison between the actual fields of the DRA (obtained using MOM) and the fields radiated by the obtained infinitesimal diploes. (a) Near fields across a line dividing the square observation plane midway. (b) Far-field.

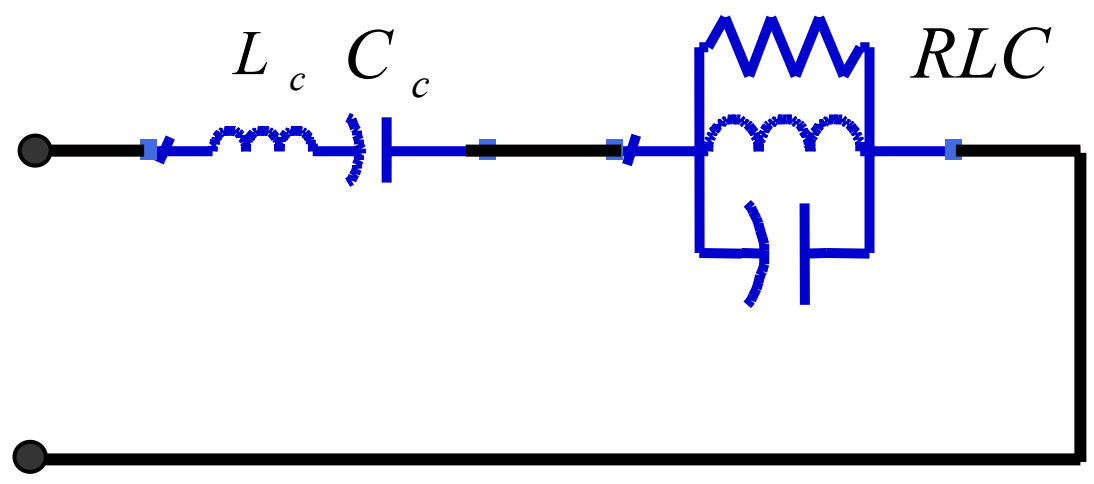

**Fig. 10** Lumped-element circuit model for the DRA including the effect coupling.

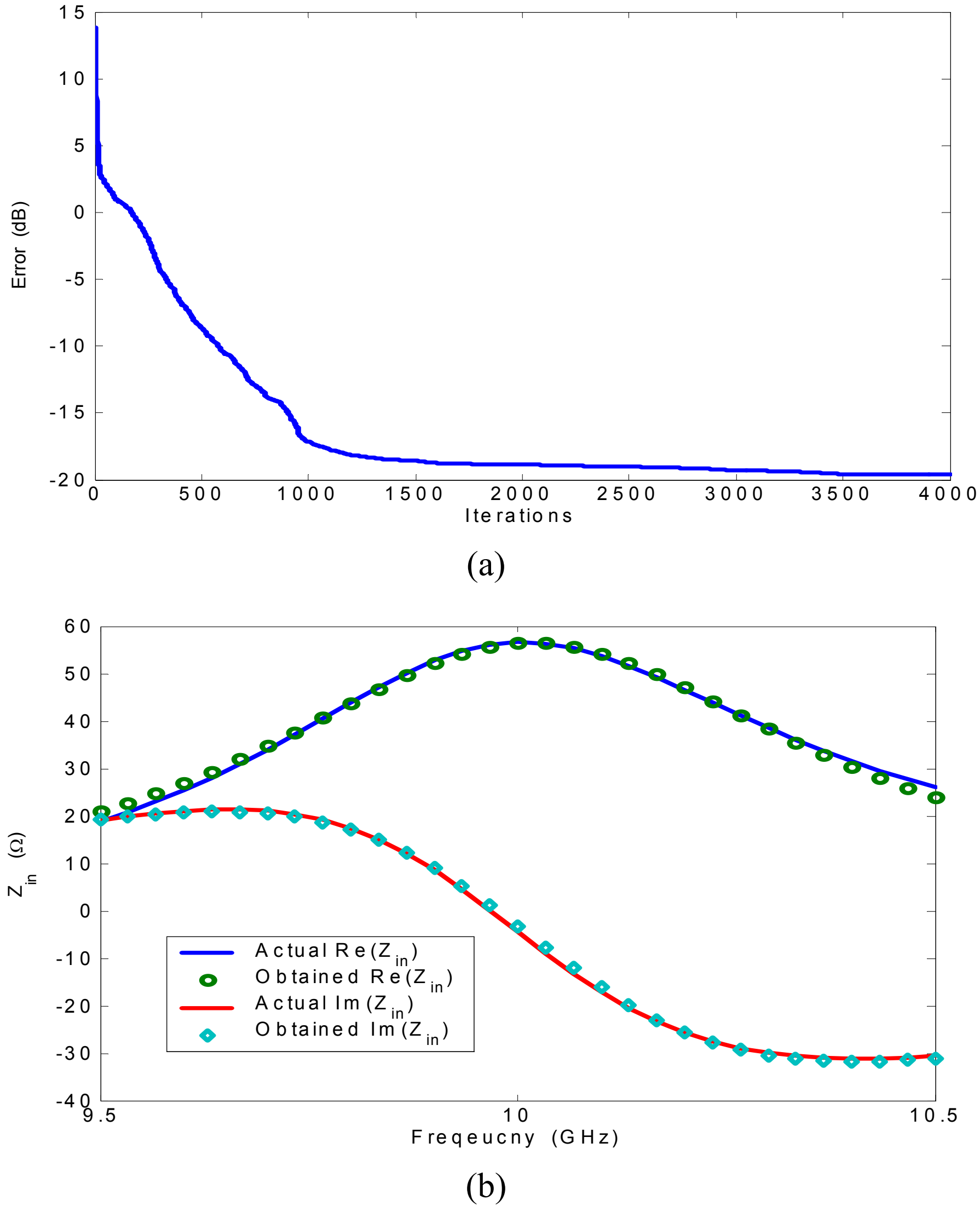

**Fig. 11** (a) Averaged convergence curve for 10 runs of the QDPSO optimization algorithm used to find equivalent circuit model for the DRA in Fig. 8. Each run consists of 4000 iterations with the dimensionality of 5. The number of particles is 30. The control parameter is $g$ = 3.0. (b) Comparison between the input impedance calculated using the circuit obtained by the QDPSO algorithm and the MOM impedance.

## References

[1] J. Kennedy and R. C. Eberhart, "Particle swarm optimization," in *Proc. IEEE Conf. Neural Networks IV*, Piscataway, NJ, 1995.

[2] G. Ciuprina, D. Ioan, and I. Munteanu, "Use of intelligent-particle swarm optimization in electromagnetics," *IEEE Trans. Magn.*, vol. 38, no.2, pp. 1037-1040, Mar. 2002.

[3] J. Robinson and Yahya Rahmat-Samii, "Particle swarm optimization in electromagnetics," *IEEE Trans. Antennas Progat.*, vol. 52, pp. 397-407, Feb. 2004.

[4] D. Boeringer and D. Werner, "Particle swarm optimization versus genetic algorithms for phased array synthesis," *IEEE Trans. Antennas Progat.*, vol. 52, no.3, pp. 771-779, Mar. 2004.

[5] Jun Sun, Bin Feng, and Wenbo Xu, "Particle swarm optimization with particles having quantum behavior," *in Proc. Cong. Evolutionary Computation*, CEC2004, vol. 1, pp. 325-331, June 2004.

[6] Said Mikki and Ahmed Kishk, "Investigation of the quantum particle swarm optimization technique for electromagnetic applications", Antennas and Propagation Society International Symposium, 2005 *IEEE* Volume 2A, pp. 45-48, 3-8 July 2005.

[7] M. Clerc and J. Kennedy, "The particle swarm: explosion, stability, and convergence in a multi-dimensional complex space", *IEEE Trans. Evolutionary Computation*, vol. 6, no. 1, pp. 58-73, Feb. 2002.

[8] F. S. Levin, *An introduction to quantum theory*, Cambridge University Press, 2002.

[9] P. A. Lindsay, *Quantum mechanics for electrical engineers*, McGraw-Hill, 1967.

[10] A. Carlisle and G. Dozier, "An off-the-shelf PSO," in *Proc. Workshop on Particle Swarm Optimization*, Indianapolis, IN, Apr. 6-7, pp. 1-6, 2001.

[11] J. R. Perez and J. Basterrechea, "Particle-swarm optimization and its application to antenna far field-pattern prediction from planar scanning," *Microwave and Optical Technology Letters*, vol. 44, no. 5, pp. 398-403, March 5 2005.

[12] Said Mikki and Ahmed A. Kishk, "Improved Particle Swarm Optimization Technique Using Hard Boundary Conditions," *Microwave and Optical Technology Letters*, vol. 46, no. 5, pp. 422-426, September 2005.

[13] Frans van den Bergh and Andries P. Engelbrecht, "A cooperative approach to particle swarm optimization," *IEEE Trans. Evolutionary Computing*, vol.8, no.3, pp. 225-239. June 2004,

[14] Asanga Ratnaweera, Saman K. Halgamuge, and Harry C. Watson, "Self-organizing hierarchical particle swarm optimizer with time-varying acceleration coefficients," *IEEE Trans. Evolutionary Computing*, vol. 8, no. 3 pp. 240-255, June 2004,.

[15] M. Wehr and G. Monich, "Detection of radiation leaks by spherically scanned field data," in *Proc. 10$^{th}$ Int. Zurich Symp. And Techn. Exhb. on EMC*, pp. 337-342, 1993.

[16] M. Wehr, A. Podubrin, and G. Monich, "Automated search for models by evolution strategy to characterize radiators," in *Proc. 11$^{th}$ Int. Zurich Symp. And Techn., Exhb. on EMC*, pp. 27-34, 1995.

[17] Joan-Rammon, Miquel Ribo, Josep-Maria Garrell, and Antonio Martin, "A genetic algorithm method for source identification and far-field radiated emissions

predicted from near-field measurements for PCB characterization", *IEEE Trans. on Electromagn. Comp.*, vol. 43, no. 4, pp. 520-530, November 2001.

[18] B. M. kolundzija, J. S. Ognjanovic, and T. K. Sarkar, *WIPL-D: Electromagnetic Modeling of Composite Metallic and Dielectric Structures, Software and User's Manual*. Reading, MA: Artech House, 2000.

[19] Taninder S. Sijher and Ahmed A. Kishk, "Antenna modeling by infinitesimal dipoles using Genetic Algorithms," *Progress In Electromagnetic Research,* PIER 52, pp. 225-254, 2005

[20] Robert E. Collin, *Foundation for Microwave Engineering*, McGraw-Hill, 1966.

[21] Darko Kajfez, *Q Factor*, Vector Fields, 1994.

[22] James Kennedy, "Probability and dynamics in the particle swarm", *Evolutionary Computation*, 2004. CEC2004. , Vol.1, pp. 340 – 347, 19-23 June 2004.